\documentclass[11pt,twoside]{article}

\usepackage{asp2006}
\usepackage{epsf}
\usepackage{lscape}
\usepackage{graphicx}

\begin{document}
\title{Model Spectra of Hot Subdwarfs for the Gaia Mission}   
\author{L.~Hindson and R.~Napiwotzki}   %%% Fill in author names
\affil{Centre for Astrophysics Research, University of Hertfordshire,
College Lane, Hatfield, AL10~9AB, UK}    
\author{U.~Heber and M.~Lemke}   %%% Fill in author names
\affil{Dr.~Remeis-Sternwarte, Universit\"at Erlangen-N\"urnberg, 
Sternwartstr.~7, 96049 Bamberg, Germany}

\begin{abstract}
The Gaia mission will provide an unprecedented 3D view of our galaxy,
it will obtain astrometric, photometric and spectrographic data for
roughly one billion stars. We are particularly interested in the
treasure chest of new data Gaia will produce for hot subdwarf B (sdB)
stars. In order for Gaia to classify sdBs and estimate parameters
model spectra covering a wide parameter range are needed. Here we
describe the construction of an extensive grid, which will be used for
this purpose. 
\end{abstract}

\section{Introduction}

Subdwarf B stars have been identified with core helium burning stars
on the extreme horizontal branch \citep{Heb1986}. The remaining
hydrogen envelope is very thin ($M_{\mathrm{env}}<0.01M_\odot$) and
they will eventually evolve more or less directly to the white dwarf
stage without an excursion to the asymptotic giant branch. However,
the origin of sdB stars is still under discussion. Proposed formation
scenarios include a late core helium flash in single star evolution
\citep{D'C.D.R1996} or close binary evolution \citep[see][for a summary
of possible scenarios]{H.P.M2002}. Radial velocity surveys of sdBs have
shown that a large fraction resides in close binary systems
\citep{M.H.M2001, N.K.L2004}. However, some tentative evidence indicates
that the binary fraction of sdBs may vary with membership in the
different galactic populations \citep{N.K.L2004, MB.M.P2006}.  Thus the
relative contributions of the evolutionary channels could be a
sensitive function of age and metallicity. 
The number of stars going through the sdB phase
could vary by a large factor for the different populations with important 
implications for, e.g., our
understanding of the UV excess in elliptical galaxies (cf.\ the contributions
of Podsiadlowski et al.\ and Yi et al.\ in these proceedings).

The astrometric satellite mission Gaia will revolutionise many fields
of stellar astrophysics, including the study of hot subdwarf
stars. Gaia will measure positions, parallaxes and proper motions of
many hot subdwarfs of spectral type sdB. This will allow us to get a
much clearer picture of space densities (and thus formation rates) and
a much improved understanding of membership in the galactic
populations.

Unambiguous classification and parameter estimates of hot subdwarfs
will be facilitated by spectra taken with a low resolution
spectrograph on board Gaia. Radial velocities of bright sdBs will be
measured with a high resolution instrument. Gaia will need a database
of spectra to perform these tasks. Here we describe the construction
of a grid of model spectra covering a large parameter space in
temperature, gravity and chemical abundances, which will be used for
the analysis of the Gaia spectra. We discuss the potential of the Gaia
spectra for parameter determinations and their limitations.

\section{The Gaia Mission}

Gaia\footnote{\tt http://www.rssd.esa.int/index.php?project=GAIA} 
is currently scheduled to be launched in December 2011 from
Kourou (French Guyana). Its final orbit will be around the Lagrangian
point L2 of the Sun--Earth system. The planned duration of the mission
is five years, during which time each object in the sky will be
observed about 70 times. The prime objective of Gaia will be high
precision astrometric measurements providing the parameters position,
proper motion and parallax. The achievable precision depends on the
brightness with expected parallax precision ranging from about
$5\,\mu$\,arcsec for $V=10$ to $200\,\mu\,$arcsec for $V=20$. 

The Photometric Instrument uses two low-resolution fused--silicia prisms
dispersing light in the along scan direction. It will provide
continuous low resolution spectra from 3200\,\AA\ to 10000\,\AA. These
will be used for classification purposes. Although the resolution is too low
to allow analysis of individual metal lines, these spectra will be
used for estimates of the
fundamental stellar parameters temperature, gravity and metallicity.

The Radial Velocity Spectrometer (RVS) will provide radial velocity
and high resolution spectral data in the narrow band 
$8470-8740$\,\AA\ (resolution $R=11500$). This range is far from optimal for
hot stars, but will nevertheless allow useful measurements for brighter
sdBs (see Fig.~2 and Sect.~3).

\section{Model Atmosphere Calculations and Simulations}

Fully metal-line blanketed LTE model atmospheres have been calculated
 using an updated version of the code described in
 \citeauthor{H.H.J1984} (\citeyear{H.H.J1984}; see also
 \citeauthor{O'T.H2006} \citeyear{O'T.H2006}). We are creating a four
 dimensional grid, varying the parameters: effective temperature:
 $20000-40000$\,K, gravity $4.5-6.5$\,dex and metallicity:
 $0.01-10\times$ scaled solar, He abundance: 1/1000--1$\times$ solar.
 Abundances in the atmospheres of sdB stars are strongly modified by
 diffusion processes, with abundance patterns often very different
 from a simple scaled solar mix \citep{E.H.N2001,O'T.H2006}. A satisfactory
 theoretical reproduction of the observations has not been achieved so
 far and the low resolution of the Photometric Instrument will prevent
 any identification of individual metal lines. The scaled solar model
 spectra are useful to get an indication of the overall metallicity.

 Model spectra are calculated with the Bamberg version of the spectrum
 synthesis programme 
LINFOR\footnote{\tt 
www.sternwarte.uni-erlangen.de/$\sim$ai26/linfit/linfor.html}
\citep{Lem1997}
 developed originally by Holweger, Steffen and Steenbock at the
 University of Kiel. Detailed line broadening tables are used for the
 hydrogen Balmer and Paschen lines and HeI and HeII lines. Metal lines
 were selected from the Kurucz line list.

\begin{figure}[!ht]
\centering
\includegraphics[height=0.9\textwidth,angle=-90]{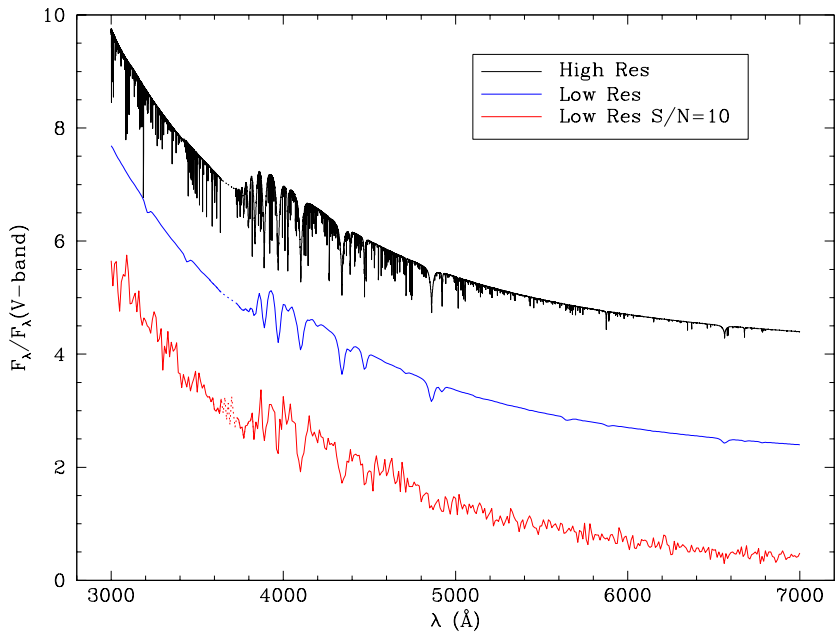}

\caption{Simple simulation of a low resolution spectrum taken by the 
Photometric Instrument for typical sdB parameters ($T_\mathrm{eff}=25000$\,K,
$\log g = 5.25$, solar abundances). 
Three stages are used to construct this spectrum. From top to bottom we
show the full resolution model spectrum, secondly the spectrum
convolved to a resolution of 25\,\AA, finally simulated noise is added at
a signal--to--noise--ratio of 10.}
\end{figure}

The model spectra were calculated on a fine wavelength grid
(0.05\,\AA).  The resolution of the Photometric Instrument spectra
will be much lower. To estimate the achievable accuracy of parameter
estimates, we performed a set of simulations. One example for typical
sdB parameters is shown in Fig.~1. These simulations demonstrate the
effect of downgrading the spectra to low resolution and photon noise.
Higher resolution spectra will be taken with the Radial Velocity
Spectrograph in the near infrared interval $8470-8740$\,\AA. This
interval was chosen, because it includes a strong CaII triplet, which
is well suited for radial velocity measurements in cool
stars. However, no strong lines are to be expected in the spectra of
sdBs (Fig.~2).  Our model calculations show a number of weaker
lines. Most moderately strong lines are He\,{\sc i} lines. Since the
photospheric He abundance of most sdBs is below the solar value,
it has to be expected that these lines are weaker in typical sdB
spectra. The strongest metal lines in this spectral range are
the two Ca\.{\sc ii} lines indicated in the plot.
Fig.~2 shows that degradation to the resolution of the
RVS will not cause a serious loss of radial velocity accuracy for
sdBs. However, a reasonable signal-to-noise ratio will be required,
limiting the usefulness of the RVS to brighter sdBs.

\begin{figure}[!ht]
\centering
\includegraphics[height=0.9\textwidth,angle=-90]{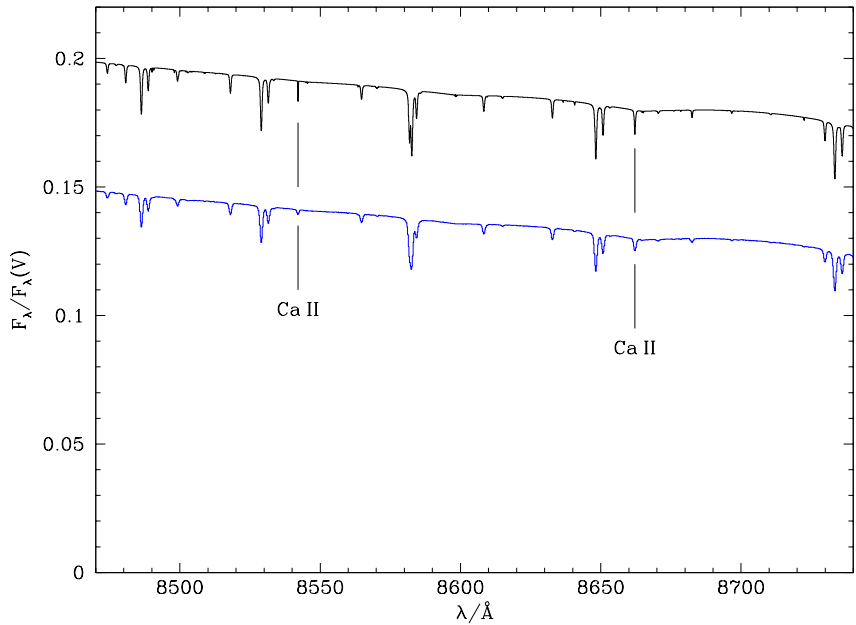}

\caption{SdB model spectrum in the Radial Velocity Spectrograph window.
Model parameters as in Fig.~1. 
Top the full resolution spectrum, below folded down to the RVS resolution.}
\end{figure}

\acknowledgements 
This project was supported by Nuffield Science Bursary URB/34488. We gratefully
acknowledge financial support of L.H. by the conference organisers.

%%% THE BIBLIOGRAPHY
%%%
%%% CONSULT SECTION 3 OF "INSTRUCTIONS FOR AUTHORS" FOR HOW TO USE NATBIB.
%%% AUTHORS ARE ENCOURAGED TO USE EITHER THE "THEBIBLIOGRAPY" ENVIRONMENT
%%% BY UNCOMMENTING (DELETING THE "%" SYMBOL) THE COMMANDS BELOW, OR BY
%%% USING THE BIBTEX ENVIRONMENT. TO FIND OUT WHICH IS APPLICABLE TO YOUR
%%% CONTRIBUTION, CONSULT THE VOLUME EDITORS FOR YOUR PROCEEDINGS.
%%%

% \bibliographystyle{apj}

% \bibliography{own,star}

\end{document}